\documentstyle[aps,epsf,openbib]{revtex}

\begin{document}
\author{G.V.Margagliotti, G.Pauli, L.Santi, S.Tessaro, A. Voronin, 
E.Zavattini} 
\title{Antiproton transfer from antiprotonic He to noble gas 
contaminants} 
\maketitle

\begin{abstract}
The state dependent quenching mechanism of metastable antiprotonic He atoms
by contaminants is suggested to explain existing experimental data. The
effect of antiproton transfer from the antiprotonic He to noble gas
contaminants is shown to play a significant role. Preliminary estimations
have been done in the framework of the coupled channels model. The obtained
results support the idea of strong dependence of quenching cross-sections on
the antiprotonic states quantum numbers and enable to explain qualitatively
existing discrepancies between experimental results, obtained for different
contaminant densities. New observable effects are predicted.
\end{abstract}

\section{Introduction.}

Recent intensive studies \cite{Iw,Yam,Obelix,Wid,Ket,Ead} of the antiproton 
($\overline{p}$) delayed annihilation in Helium allowed the discovery of
unique features of antiprotonic He systems. Theoretical explanation of
abnormal lifetime ($\tau \sim 10^{-6}$s) of certain antiproton fraction in
He requires the existence of antiprotonic He systems which have radiative
and Auger de-excitation lifetime of the order of $10^{-6}$ s and are very
stable with respect to the thermal collisions with surrounding He atoms.

It was first suggested by Condo \cite{C} and validated by further detailed
studies \cite{Rus,Sh,Oh,Mo,Kor,VD} that highly excited circular (or near
circular) states of antiprotonic He atoms (He$^{+}\overline{p}$)$_{N,L}$
(here N is principal quantum number and L is the angular momentum quantum
number of the antiprotonic state) should have extremely small (in atomic
scale) Auger de-excitation rates ($\lambda _{A}\ll 10^{6}$ s$^{-1}$for $N,
L{>%
}36$). The lifetime of such highly excited antiprotonic atoms, if they are
isolated from collisions with surrounding medium, is determined by the
radiative transitions and is of the order of $10^{-6}$ s. This fact
motivates for treating antiprotonic He atom as a system responsible for
delayed annihilation. Meanwhile, the complete understanding of the problem
can be obtained only carefully analyzing the effects of the collisions of
such a system with surrounding medium \cite{Val,GK,Vor}.

It has been shown in \cite{Vor} that different states of metastable
antiprotonic atoms could be affected by several collisional quenching
mechanisms, which are Stark transitions to nonstable states, collisionally
induced Auger decay and rearrangement processes, like short living molecular
ion formation.

In this paper we analyze those rearrangement collisions which result in the
antiproton transfer from high metastable states of the antiprotonic He (with
principal quantum number $N\geq 40$) to noble gas contaminant atoms. Some
remarkable features enable to distinguish this process among others. First,
one can expect that such transfer of antiproton to nonhelium atom will
result in the fast Auger de-excitation \cite{VD} and following annihilation
of antiproton on the contaminant nucleus, which can be checked
experimentally. Second, it is reasonable to expect classical character of
antiprotonic transfer, thus the corresponding reaction cross-sections should
be of the order of geometrical atomic cross-sections at least for certain
states of antiprotonic He. We will show that taking into account antiproton
transfer mechanism, it is possible to obtain qualitative explanation of
existing experimental data on contaminant quenching, in particular about the
apparent discrepancy between experimental results on noble gas contaminant
quenching obtained by different experimental groups.

\section{Experimental overview.}

Existing experimental results on quenching of antiprotonic atom metastable
states by noble gas contaminants were obtained by OBELIX collaboration (PS
201) \cite{Obelix} and CERN group (PS205) \cite{Wid}. The averaged over
different metastable states quenching cross-sections obtained by the two
groups are shown in Table 1, with their corresponding ratios.

\begin{center}
$%
\mathrel{\mathop{%
\begin{tabular}{|c|c|c|c|}
\hline
Contaminant & $\sigma _{quench}^{1}$ & $\sigma _{quench}^{2}$ & 
$\sigma _{quench}^{1}/\sigma _{quench}^{2}$ \\ \hline
Ne & 2*10$^{-17}$cm$^{2}$ & 10$^{-20}$cm$^{2}$ & 2*10$^{3}$ \\ \hline
Ar & 7*10$^{-17}$cm$^{2}$ & 4*10$^{-20}$cm$^{2}$ & 1.7*10$^{3}$ \\ \hline
Xe & 2*10$^{-16}$cm$^{2}$ & 3*10$^{-18}$cm$^{2}$ & 0.6*10$^{2}$ \\ \hline
\end{tabular}
}\limits_{\text{{\bf Table I}. {\rm Quenching cross-section for different 
contaminants obtained by OBELIX (1) and PS205 (2) groups}.}}}%
$
\end{center}

\bigskip

A dramatic difference between the results of the two groups can be seen,
specially for Ne and Ar. We must notice that the mentioned experiments have
been done under different conditions, among which we outline the different
contaminant concentration. In OBELIX experiment the same quenching rate was
observed with concentrations of noble gas contaminant several orders of
magnitude less than in PS205 experiment. Unfortunately, there are no data
obtained in the overlapping ranges of contaminant densities.

Such a discrepancy suggests the idea that quenching effects are very
different for different metastable antiprotonic states in the sense that the
lifetime of certain metastable states is measurable for small contaminant
concentrations only, while for big enough concentrations it becomes too
short to be distinguished from the prompt peak. Thus the averaged over
antiprotonic states quenching cross-section becomes a function of
contaminant density.

We will show that estimation of the antiproton transfer rates supports for
such an explanation of the difference in extracted quenching cross-section
values.

\section{Rearrangement collisions.}

The correct description of inelastic collisions of an antiprotonic atom with
medium atoms requires taking into account simultaneously all possible
reaction channels. We will show, however, that it is meaningful to
distinguish the antiproton transfer mechanism among other processes and
treat it separately. Let us first make some qualitative remarks. We will be
interested in the reaction:

\begin{equation}
(He^{+}\overline{p})_{N,L}+ \text{A} \longrightarrow \left[ He^{+}e\overline{%
p} \text{A}^{+} \right] \longrightarrow \left[ He( \text{A}^{++}\overline{p}%
)_{N^{\prime},L^{\prime }} \right] +e  \label{AT}
\end{equation}

where A reads essentially for a contaminant atom present in the surrounding
medium.

In this reaction the exchange of electron and antiproton between He and
contaminant atom A takes place, which results in a virtual formation of the
molecular system $\left[ He^{+}e\overline{p} \text{A}^{+}\right] $. The
energy excess is then transferred to an Auger electron of the contaminant
atom, while the molecular ion $\left[ He( \text{A}^{++}\overline{p}%
)_{N^{\prime },L^{\prime }}\right] $ is formed in the final state. We notice
that the more simple reaction:

\begin{equation}
(He^{+}\overline{p})_{N,L}+ \text{A} \longrightarrow He+( \text{A}^{+}%
\overline{p} )_{N^{\prime },L^{\prime }}  \label{AT1}
\end{equation}

occurs with significant probability only in the resonance case, i.e. when
bounding energy of $(He^{+}\overline{p})_{N,L}$ is equal to that of $(\text{A%
}^{+}\overline{p})_{N^{\prime },L^{\prime }}$. In nonresonance case an
energy excess (which characteristic value in mentioned reactions is about
0.1 eV) is transferred to atomic nuclei relative motion and no term crossing
takes place in this case. Thus reaction probability turns to be
exponentially small. An obvious exception is antiprotonic Helium collision
with He atoms of surrounding medium:

\begin{equation}
(He^{+}\overline{p})_{N,L}+He\longrightarrow He+(He^{+}\overline{p})_{N,L}
\label{AT2}
\end{equation}

As it follows from our calculations, reaction (\ref{AT2}) takes place for
antiprotonic states with principal quantum number $N>42$. Such states are
already quenched within short time by Stark de-excitation collisions \cite
{GK,Vor} and can not be observed within the delayed component. In the same
time we will show that reaction (\ref{AT}) for Ne and Ar contaminants affect
certain states belonging to the observed delayed fraction.

The amplitude of reaction (\ref{AT}) is determined by the overlapping of the
antiprotonic wave function of $(He^{+}\overline{p})_{N,L}$ and the
antiprotonic wave function of (A$^{++}\overline{p})_{N^{\prime },L^{\prime
}} $. We will show later, that there is a repulsive barrier in the effective
interaction of antiprotonic He and noble gas atom, which prevents close
collisions. Clearly antiproton transfer has a chance only if the interatomic
separation becomes small enough during the collision, to ensure overlapping
of the wave function of antiproton centered on He, and that of medium atoms.
This last condition determines the reaction probability dependence on
antiprotonic quantum numbers. We will show that for the interaction of
antiprotonic He with Ne and Ar, antiproton transfer takes place for states
with $N\geq 40.$

\subsection{\protect\bigskip Formalism.}

We search the wave function of the system ($He^{+}\overline{p}$)$_{N,L}+$ A
in the form:

\begin{eqnarray}
\Phi &=&\sum_{\alpha \beta \gamma }^{{}}\chi _{\gamma }^{\overline{p}}%
\widehat{P}\left[ \varphi _{\alpha }^{e}\Psi _{\beta }^{e}\right] F_{\left\{
\alpha \beta \gamma \right\} }  \nonumber \\
&&+\sum_{\alpha \beta \gamma \delta }\widetilde{\chi }_{\gamma }^{\overline{p%
}}\widehat{P}\left[ g_{\left\{ \alpha \beta \gamma \delta \right\}
}^{e}\varphi _{\alpha }^{e}\widetilde{\Psi }_{\beta }^{+e}\right] Y_{\delta }
\label{FS}
\end{eqnarray}

The functions $\varphi _{\alpha }^{e}$, $\Psi _{\beta }^{e}$, $\chi _{\gamma
}^{\overline{p}}$ are the eigenfunctions of the electron in the field of He
nuclei, contaminant atom electron wave function, and $\overline{p}$ \ wave
function in the field of He nuclei, screened by electron in the ground state
respectively. The functions $\widetilde{\Psi }_{\beta }^{+e}$, $\widetilde{%
\chi }_{\gamma }^{\overline{p}}$, $Y_{\delta }$ are the electron wave
function of contaminant ion (with charge +1), the $\overline{p}$ \ wave
function centered on contaminant nuclei, the wave function of the nuclei
relative motion in bound (molecular) state respectively. $\widehat{P}$ is
the permutation operator, which antisymmetrizes the total electronic wave
function. The expansion coefficient $F_{\left\{ \alpha \beta \gamma \right\}
}$ has the sense of the nuclei relative motion wave function in the
scattering state, while $g_{\left\{ \alpha \beta \gamma \delta \right\}
}^{e} $ can be interpreted as Auger electron wave function. $F_{\left\{
\alpha \beta \gamma \right\} }$and $g_{\left\{ \alpha \beta \gamma \delta
\right\} }^{e}$ include reaction amplitudes to be find.

The mentioned form of the wave function enables to take into account
physically important effects of exchange of the electrons and antiproton
between nuclei, as well as the antisymmetrization of the electronic
wave-function. We obtain the coupled equations system for functions $%
F_{\left\{ \alpha \beta \gamma \right\} }$, $g_{\left\{ \alpha \beta \gamma
\delta \right\} }^{e}$ by substituting expansions (\ref{FS}) in the
Shrodinger equation for the interacting systems.

For the purpose of qualitative estimations of the rate of the exchange
mechanism, we have truncated the mentioned equation system to only few
coupled equations.

\subsection{ Interaction potential.}

The coupled equation system for $F_{\left\{ \alpha \beta \gamma \right\} }$, 
$g_{\left\{ \alpha \beta \gamma \delta \right\} }^{e}$ can be transformed
into the one-channel Shrodinger equation for the relative nucleus motion in
the elastic channel $F_{\left\{ \alpha _{0}\beta _{0}\gamma _{0}\right\}
}\equiv F_{\left\{ \xi _{0}\right\} }$ :

\begin{equation}
\left( \widehat{T}_{_{A}}+\widehat{V}_{AHe}^{\left\{ \xi _{0}\right\} }-i%
\widehat{W}_{AHe}^{\left\{ \xi _{0}\right\} }-E^{\left\{ \xi _{0}\right\}
}\right) F_{\left\{ \xi _{0}\right\} }=0  \label{Eff}
\end{equation}

\begin{figure}[tbp]
\vspace{+4.8cm} \epsfxsize=12cm\epsfysize=12cm
\par
\begin{center}
\mbox{\epsffile{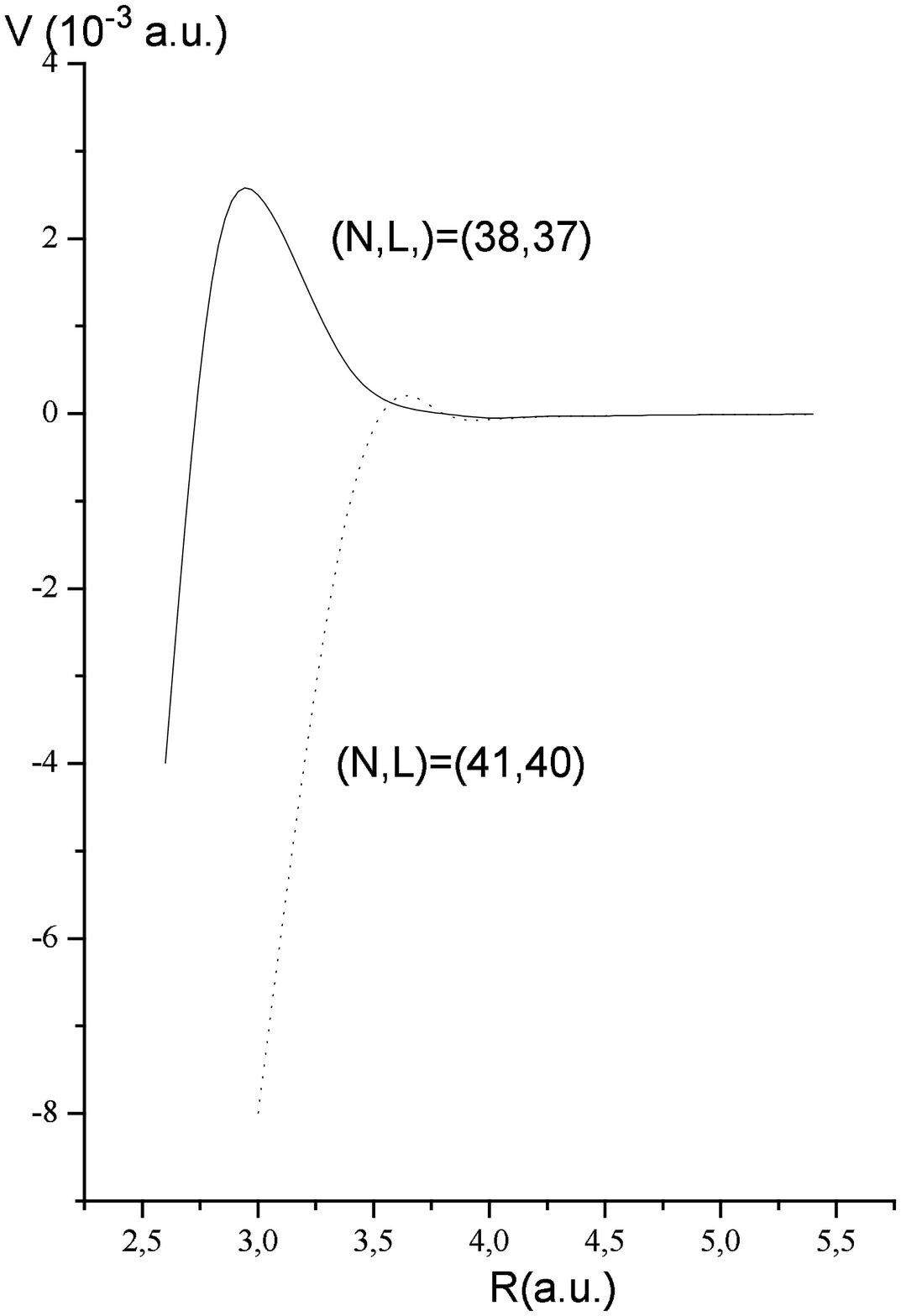}}
\end{center}
\caption{State dependent antiprotonic atom-media He atom repulsive barrier }
\label{Fig1}
\end{figure}

Such an equation includes a complex nonlocal interaction term $\widehat{V}%
_{AHe}^{\left\{ \xi _{0}\right\} }-i\widehat{W}_{AHe}^{\left\{ \xi
_{0}\right\} }$. This interaction describes elastic scattering and
absorption into inelastic channels and depends on quantum numbers $\left\{
\xi _{0}\right\} $. It turns out that leading terms of the real part $%
\widehat{V}_{AHe}^{\left\{ \xi _{0}\right\} }$ of such effective interaction
have local form and can be interpreted as antiprotonic atom-media atom
potential in given state. We should mention that both local and nonlocal
terms in $\widehat{V}_{AHe}^{\left\{ \xi _{0}\right\} }-i\widehat{W}%
_{AHe}^{\left\{ \xi _{0}\right\} }$ are important for reaction rates
calculation, nevertheless the analysis of local real terms alone turns to be
very useful . Such a potential for (He$^{+}\overline{p}$)$_{N,L}-$He
interaction is shown on Fig.\ref{Fig1}. Important features of this potential
are the following:

\begin{enumerate}
\item  There is a repulsive barrier between ($He^{+}\overline{p}$)$_{N,L}$
and He at internuclear distance 3 au$<R_{{}}<5.5$ au (Fig.\ref{Fig1}).
 The height of this barrier strongly depends on N,L (see
also \cite{Val}). Its height is about 0.2 eV for N=38, L=37 and is
negligible for $N>42$. Such a barrier appears as a result of
antisymmetrization of 3-electron wave function of interacting ($He^{+}%
\overline{p}$)$_{N,L}-$He atoms and represents an effect of Pauli repulsion.
The minimum classically allowed interatomic separation distance, which is
determined by this repulsive part of effective interaction, plays an
important role for determination of quenching reaction rates. For the
antiprotonic state with N=39 it was found to be $R_{c}=5.3$ au. The
repulsive barrier appears also in ($He^{+}\overline{p}$)$_{N,L}-$Ne and ($%
He^{+}\overline{p}$)$_{N,L}-$Ar effective interaction. As it follows from
our calculations Ne and Ar can penetrate to short enough distances during
the collision with antiprotonic He and this is a crucial point for
estimation of antiprotonic transfer reactions.

\item  At the internuclear distances $R_{{}}$ from 1 au. to 3 au., the ($%
He^{+}\overline{p}$)$_{N,L}-$He potential is attractive. This attraction is
mainly due to the $\overline{p}$ exchange between the two nuclei. The range
of the attractive part is determined by the overlapping of antiprotonic
states, centered on the two nuclei; it vanishes rapidly as soon as the
internuclear distance becomes grater than two mean radii of antiprotonic
state with quantum numbers N and L. This part of interaction is important
for antiprotonic transfer reactions.

\item  At large internuclear distances there is a weak polarization
attraction between ($He^{+}\overline{p}$)$_{N,L}$ and contaminant atom A: 
\[
\widehat{V}_{AHe}^{\left\{ \xi _{0}\right\} }\rightarrow -\frac{%
C_{AHe}^{\left\{ \xi _{0}\right\} }}{R_{{}}^{6}}
\]
\end{enumerate}

\bigskip

This long range attractive interaction radically enhances inelastic
cross-sections, specially in case of low temperatures (T%
\mbox{$<$}%
300K). The constant $C_{AHe}^{\left\{ \xi _{0}\right\} }$ depends on
contaminant. This last statement is important for understanding the
difference in quenching effect of noble gas contaminants.

We should mention that imaginary part $\widehat{W}_{AHe}^{\left\{ \xi
_{0}\right\} }$ of the effective interaction is localized mainly at
internuclear distances $R_{{}}\leq $3 au. Thus the repulsive barrier between
($He^{+}\overline{p}$)$_{N,L}$ and He for $N<42$ prevents close collisions,
which may result in intensive inelastic transitions and quenching of
metastable antiprotonic states.

\subsection{Quenching cross-sections.}

In this subsection we present the estimation of the antiproton transfer
cross-sections for different states of ($He^{+}\overline{p}$)$_{N,L}$.

We found strong dependence of quenching cross-sections on the principal
quantum number of the antiprotonic atoms. In particular, the corresponding
cross-sections of antiproton transfer for the states with $40\leq N<42$ are:

\[
\sigma _{40\leq N<42}^{Ne,Ar}\approx 10^{-17}\mbox{cm}^{2} 
\]

Mentioned states become short-living ($\tau \approx 10^{-7}$ s) in the
presence of noble gas contaminants with density $\rho \approx 10^{18}$ cm$%
^{-3}$.

\begin{figure}[tbp]
\vspace{+4.8cm} \epsfxsize=12cm\epsfysize=12cm
\par
\begin{center}
\mbox{\epsffile{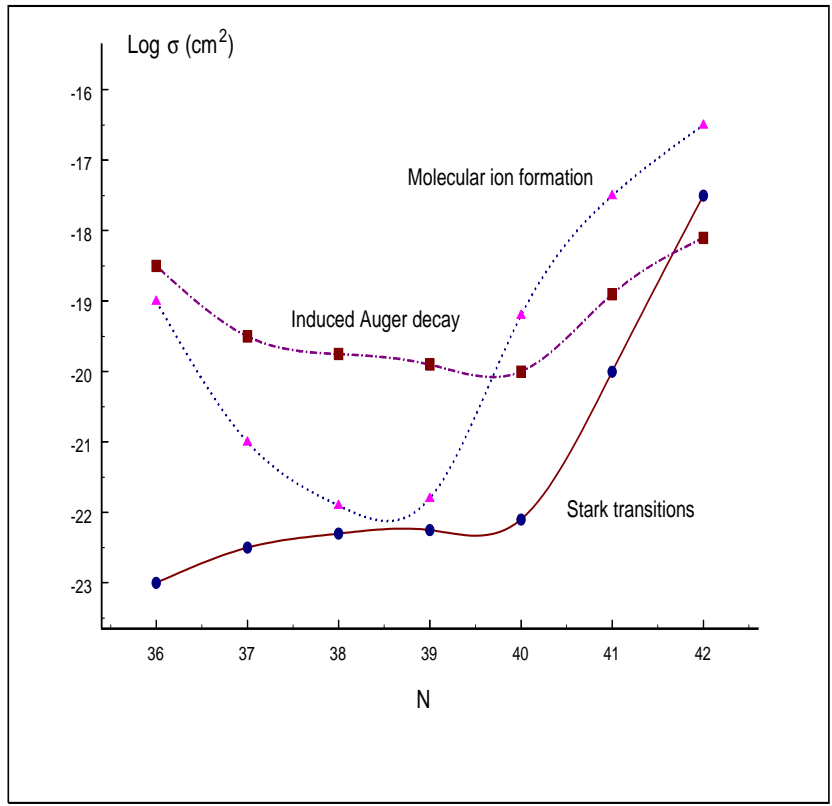}}
\end{center}

\caption{Ne contaminant quenching cross-sections  as a function of
antiprotonic atom state principal quantum number}
\label{Fig2}
\end{figure}

In the same time the antiproton transfer probability for states with $N\leq
39$ is negligible(see Fig.\ref{Fig2}). Such a ''threshold'' behavior of the
transfer cross-section as a function of principal quantum number is clear
from the following qualitative argument. As it follows from the properties
of effective interatomic interaction, the less is N of given antiprotonic
state the higher is the repulsive barrier and the bigger is interatomic
separation during the collision. On the other hand the less is N, the less
is the overlapping of the antiprotonic functions centered on He and
contaminant atom, respectively for given interatomic separation. We have
found that N=40 plays a role of critical number for antiproton transfer from
He to Ne and Ar.

\subsection{Experimental check.}

The existing experimental data on noble gas contaminant quenching can now be
explained in terms of state dependent quenching mechanism.

The antiprotonic transfer reactions affect the population of the states with 
$40\leq N <42$. These states are long living in the absence of contaminants.
Rather small concentration of contaminant gases, like those used by OBELIX ($%
\rho \approx 10^{17}\mbox{cm}^{-3}$), can produce measurable effects ($%
\lambda_{quench} \approx 10^{6}\mbox{s}^{-1} $). The averaged over states
quenching cross-section, derived from OBELIX data, correspond to the
antiproton transfer reaction cross-sections.

Much higher concentration of contaminant Ne or Ar( $\rho \approx 10^{20}%
\mbox{cm}^{-3}$), used in PS205 experiment, produce quenching rates $\lambda
_{quench}\approx 10^{8}\mbox{s}^{-1}$, which make impossible to distinguish
such states from the prompt peak. On the other hand the states with $N<40$
are not quenched by antiproton transfer reactions. The main contaminant
quenching mechanism for these states is induced Auger de-excitation \cite
{GK,Vor} (see Fig.\ref{Fig2}). The corresponding quenching cross-sections
are \cite{Vor}:

\[
\sigma _{N<40}^{Ne,Ar}\approx 10^{-19}\mbox{cm}^{2} 
\]

The concentration of noble gas contaminants required to produce measurable
quenching of these states is $\rho \approx 10^{20}\mbox{cm}^{-3}$,
corresponding to those used in PS205 experiment. Thus it may be expected
that the results obtained by PS205 experiment refer to the quenching of
states with $N<40$. This fact enables to understand qualitatively the
discrepancy between the results of the two experimental groups (OBELIX and
PS205) for Ne and Ar contaminant quenching \cite{Obelix,Wid}. The difference
in the contaminant densities ($\rho \approx 10^{17}$ cm$^{-3}$ for PS205,
and $\rho \approx 10^{20}$ cm$^{-3}$ for OBELIX) correspond to the
difference between the extracted values of the average quenching
cross-sections.

Some experiments can be suggested to clarify the situation.

First it seems reasonable to obtain results for the whole range of
contaminant densities to check if asymptotic behaviors of quenching rates,
which are different in the two experiments, match at the intermediate
densities.

The direct check of the antiproton transfer reactions could be the
observation  of heavy fragments produced by antiproton annihilation on
contaminant nuclei among the delayed events.

The laser spectroscopy methods, similar to those applied for observing $%
H_{2} $ assisted resonances \cite{Ket}, seem to be also useful to study
noble gas contaminant quenching. In fact, inducing laser transition from
state with N=39 to states with $N\geq 40$ in the presence of Ne or Ar at
densities $\rho \approx 10^{18}\mbox{cm}^{-3}$, one should observe resonance
in annihilation events having width proportional to the contaminant
concentration.

\section{Conclusion.}

We have found that the approach, in which the state dependence of quenching
rates is taken into account, enables to explain existing experimental data.
The mentioned above antiproton transfer reaction rates indeed have very
sharp dependence on antiprotonic Helium state quantum numbers. The
theoretical model suggested here, is based on the following statements.

\begin{enumerate}
\item  For the states with $N<42$ and Auger transition multipolarity $\Delta
l>3$ there is a repulsive barrier which prevents from close collisions in
''antiprotonic atom-medium atoms'' interaction and therefore plays a
stabilizing role. The physical reason of such barrier is Pauli repulsion of
the saturated electronic shell of noble gases and the electron of
antiprotonic He. The mentioned barrier determines the minimum separation
between atoms during collision, on which quenching reaction rates critically
depend.

\item  The leading contaminant quenching mechanism of metastable
antiprotonic He states with $40\leq N<42$ is antiproton transfer reaction,
followed by fast antiproton annihilation on the contaminant nucleus. The
cross-section of this type of reactions is estimated to be:

\[
\sigma _{40\leq N<42}^{Ne,Ar}\approx 10^{-17}\mbox{cm}^{2}
\]

In the same time the noble gas contaminant quenching of states with $N<40$
is two orders of magnitude less.

\item  The evolution of the antiprotonic atoms passes through the stage of
molecular ion formation, especially in the presence of noble gas
contaminant. This fact was first pointed out by E. Zavattini \cite{ZB}. In
the present work we studied short living antiprotonic molecular ion
formation. In the same time the problem of possible existence of long living
states of antiprotonic molecular ion remains an open question.
\end{enumerate}

Experimental test of the above presented theoretical results may include
direct observation of heavy fragments among the delayed annihilation events,
produced by antiproton annihilation on contaminant nuclei as well as laser
induced transitions from states with $N<40$ to states with $N\geq 40$ in the
presence of Ne or Ar at concentrations $\rho \approx 10^{18}\mbox{cm}^{-3}$.

\section{Acknowledgment.}

One of the authors (AV) would like to thank P. Valiron, J. Carbonell and G.
Korenman for useful discussions and express his special acknowledgment to
Italian Istituto Nazionale di Fisica Nucleare (INFN) for financial support.


\newpage 


\newpage 


\end{document}